\documentclass{article}
\usepackage[utf8]{inputenc}
\usepackage{amssymb}
\usepackage{color}
\usepackage{xcolor}
\usepackage{graphicx}
\usepackage{graphics}
\usepackage[colorlinks=true,linkcolor=red,filecolor=green,citecolor=red]{hyperref}
\usepackage{cleveref}
\usepackage{url,lineno,microtype,subcaption,comment}

\begin{document}
\onecolumn

\title{Seller-buyer networks in NFT art \\ are driven by preferential ties} 

\author{Giovanni Colavizza\thanks{University of Amsterdam, The Netherlands. \url{g.colavizza@uva.nl}.}}
\date{}



\maketitle

\begin{abstract}


Non-Fungible Tokens (NFTs) have recently surged to mainstream attention by allowing the exchange of digital assets via blockchains. NFTs have also been adopted by artists to sell digital art. One of the promises of NFTs is broadening participation to the art market, a traditionally closed and opaque system, to sustain a wider and more diverse set of artists and collectors. A key sign of this effect would be the disappearance or at least reduction in importance of seller-buyer preferential ties, whereby the success of an artist is strongly dependent on the patronage of a single collector. We investigate NFT art seller-buyer networks considering several galleries and a large set of nearly 40,000 sales for over 230M USD in total volume. We find that NFT art is a highly concentrated market driven by few successful sellers and even fewer systematic buyers. High concentration is present in both the number of sales and, even more strongly, in their priced volume. Furthermore, we show that, while a broader-participation market was present in the early phase of NFT art adoption, preferential ties have dominated during market growth, peak and recent decline. We consistently find that the top buyer accounts on average for over 80\% of buys for a given seller. Similar trends apply to buyers and their top seller. We conclude that NFT art constitutes, at the present, a highly concentrated market driven by preferential seller-buyer ties.

\end{abstract}

\section{Introduction}

More than half the global visual art market is made of sales above \$1M, largely by so-called `ultra-high-net-worth individuals'. The middle-tier market is thinning, with allegedly 60\% of galleries operating at a loss~\cite{deresiewicz_death_2020}. While artists start their careers with high expectations, the vast majority never makes a living from their artistic practice~\cite{plattner_most_1998}. In the overall `good' year of 2018, fifty artists accounted for 64\% of global auction sales, the top-500 artists accounted for 89\% of total sales~\cite{artprice_contemporary_2019}. While artistic careers have largely resisted systematic empirical study, it is known that success in the arts is influenced by a multifaceted mixture of early and self-reinforcing access to opportunity~\cite{fraiberger_quantifying_2018}, tightly knit social networks~\cite{giuffre_sandpiles_1999}, and even sheer luck~\cite{janosov_success_2020}. Why is it so? The traditional art market deals in heterogeneous, mostly unique goods (artworks), bought by collectors with equally heterogeneous tastes~\cite{towse_art_2011}. As a result, uncertainty reigns. Both sellers and buyers are at a disadvantage, whereas intermediaries retain substantial brokering power~\cite{gertsberg_asymmetries_2019}. These include dealers and galleries, auction houses, experts, curators, and cultural institutions. They match offer and demand by reducing information asymmetries for a margin~\cite{akerlof_market_1970}. Seemingly, not even the Internet has disrupted the art market `stasis'~\cite{velthuis_contemporary_2012}. 

Such feats of the traditional art market go a long way in explaining why many artists and collectors enthusiastically embraced \textbf{Non-Fungible Tokens (NFTs): certificates of authenticity registered on public blockchains}. The NFT blockchain technology brings with it the promise to enable radically different, actor-first markets. 

NFTs constitute a new asset class solving a significant problem: digital ownership. Markets cannot operate without property rights, and digital assets have been most elusive in this respect~\cite{kaczynski_how_2021}. NFTs establish at once an immutable chain of provenance and digital scarcity for an object. What is more, NFTs are programmable, therefore they can be bestowed with deterministic behavior when they are created or afterwards, expanding their purpose in an open-ended way. For example, NFTs can be programmed to yield royalties to the original creator over any transaction in the market of re-sales~\cite{tapscott_blockchain_2017}. From a survey among NFT artists, this emerged as the most appreciated innovation~\cite{braidotti_crypto_2021}. In 2021, NFTs have rapidly gone mainstream and seen widespread adoption~\cite{murphy_how_2021,nadini_mapping_2021}. 

NFTs promise solutions to many open challenges: they bring full transparency since all transactions are openly stored on a blockchain, providing immediate access to liquidity and better price discovery. They support socio-technical decentralization, shifting the power balance from intermediaries to actors (artists and collectors), supporting novel ownership and governance solutions to emerge~\cite{evans_cryptokitties_2019}. The effects are visible: leading NFT marketplaces SuperRare and ArtBlocks have gallery fees of 10-15\% on primary sales, and 0-3\% on re-sales; compare with an average of 30-70\% for galleries and 20\% for auction houses in the traditional art market~\cite{towse_handbook_2020}. In 2021 alone SuperRare paid about \$10M in artist royalties on re-sales (source \textit{cryptoart.io}); compare to zero paid by traditional auction houses. This is why many view NFTs as building blocks in the realization of an open and sustainable Internet~\cite{nguyen_value_2021}. 

The art market is comparable to other sectors where intermediaries hold positional advantages. Examples include centralized financial services, social media platforms and scientific publishing. The arts and collectibles markets have been among the first to adopt NFTs, constituting a relatively mature stage of NFTs adoption~\cite{franceschet_crypto_2021}. This is evidenced by the move of traditional art market players into NFTs and the de-facto entry of NFT art into the contemporary art asset class~\cite{kinsella_boosted_2021}. The expansion of NFTs in many other sectors is also unfolding, for example online gaming~\cite{fowler_tokenfication_2021}. 

There remains considerable skepticism around NFTs, and blockchains more broadly. The main challenges stem from: i) the absence of a legal and regulatory framework around NFTs, making their actual ownership rights unclear at best~\cite{savelyev_copyright_2018}; ii) technical risks due to persisting platform centralization~\cite{marlinspike_my_2022} and the environmental costs of mining~\cite{kay_hidden_2021}; iii) the possible negative impact of rampant financial speculation and, sometimes, illicit practices~\cite{murphy_how_2021}. What is more, blockchain technology, and NFTs within it, are subject to cycles of booms and busts which might significantly influence the pricing of assets and the long-term development of products. 

In fact, we know very little about the potential and actual impact of NFTs. Empirical research on NFTs’ social, economic and cultural implications is just beginning~\cite{nadini_mapping_2021,vasan_quantifying_2022}. We know little about how NFTs are used in practice, whether they are delivering social and economic impact to a broad set of actors, what are the costs and risks involved. As a consequence, the scientific and public debate often contrasts opposite viewpoints, with harsh critiques facing radical supporters~\cite{zeilinger_digital_2018,lotti_art_2019,joselit_nfts_2021}. The systematic analysis of data, made eminently possible by its availability through public blockchains, is essential to ground such discussions in facts. 

This contribution focuses on the key relationship in art markets: the one between sellers and buyers, thus most often artists and collectors. \textbf{Our goal is to provide a quantitative overview of the NFT art market from the 2020-2021 boom to the 2022 bust, through the lenses of seller-buyer networks.} The main question we ask is whether NFT art is collected from a more diverse set of buyers than fine arts, or instead whether strong preferential ties constitute a persistent feature of art markets even when supported by public blockchains. Previous work has shown that NFT traders tend to specialize within NFT segments and collections, and that the market is highly concentrated~\cite{nadini_mapping_2021}. When  considering NFT art, preliminary evidence points to significant similarities with the traditional art market in terms of systematic seller-buyer ties, sticky reputation effects, and high degrees of collector specialization~\cite{vasan_quantifying_2022}.

\section{Related work}

\subsection{Blockchains}

A blockchain is an append-only, distributed database, that users store and update via a decentralized protocol~\cite{narayanan_bitcoin_2016}. Cryptography is used to avoid tampering and falsification with the data a blockchain stores. The origins of the blockchain date back to the work of ~\cite{haber_how_1991} on how to create tamper-proof data. The first, popular implementation of a blockchain is bitcoin~\cite{nakamoto_bitcoin_2008}: a financial blockchain exclusively devoted to tracing transactions related to minting and exchanging the bitcoin crypto currency. Bitcoin introduced the idea of `proof of work' as its decentralized protocol (also called `consensus mechanism'). The paramount problem blockchain technology solves is of great social importance: reaching consensus on authoritative records in the absence of centralized institutions and trusted third parties. The use of blockchains has been growing rapidly over the past few years, attracting widespread attention~\cite{casino_systematic_2019,xu_systematic_2019}. Blockchain technology is creating a space to experiment with novel forms of decentralized organizations, markets and coordination systems~\cite{wright_decentralized_2015,olnes_blockchain_2017}. A key development, in this respect, has been the introduction of programmable blockchains, able to record a variety of user activities via `smart contracts'. The most well-known example is Ethereum~\cite{buterin_ethereum_2013}. Despite the growth in interest and adoption of blockchain technology, there is still limited evidence-based research on its societal, economic, and cultural impact, as well as long-term potential.

The potential of blockchain technology applications in the arts has been much anticipated and debated, primarily theoretically or relying on circumscribed qualitative studies~\cite{sidorova_cyber_2019}. Practitioners consider the arts ``one of the least discussed applications for blockchain, yet one where the technology may hit hardest''~\cite{macdonald-korth_art_2018}. \cite{whitaker_primer_2019} identifies at least three transformations which are made possible by the affordances of blockchain technology: 1) ``blockchain blurs the for-profit/nonprofit distinction in the arts because the decentralized structure shifts responsibility for infrastructure away from trusted central authorities'' such as public property registers, or institutional intermediaries (e.g., galleries and museums); 2) ``blockchain changes the ownership structure of art by creating fractional ownership of artworks and scarcity for digital works, [..] these potential shared-value structures extend to resale royalties and copyright''; 3) ``blockchain’s shared value structures generalize to new models of supporting the arts itself.'' Indeed, blockchain technology has been used in a variety of applications, from enabling novel galleries and marketplaces~\cite{wang_non-fungible_2021} to tracking looted or contested cultural heritage~\cite{whitaker_art_2021}.

Artists often see blockchains as a possible means to bring about a re-centering of the market around and in support of their creative processes, with a sustainable and inclusive outlook~\cite{bollier_blockchain_2015,catlow_artists_2017}, where ``artists set the terms of their market participation''~\cite{cunningham_web3_2019}.

\subsection{Non-Fungible Tokens (NFTs) and the arts}

The key blockchain innovation underpinning this vision are Non-Fungible Tokens (NFTs). Art registered as an NFT is called rare digital art, crypto art, blockchain art or NFT art~\cite{bailey_what_2018,franceschet_crypto_2021} (we use NFT art throughout). NFT art may be considered a form of digital art~\cite{paul_companion_2016}: NFT artworks often use digital media and do not share a unique set of aesthetics. Furthermore, a share of NFT art is born-analog and has been digitized in order to register it as an NFT. Many of the innovations NFTs are bringing to the arts have antecedents in early blockchain projects such as Monegraph~\cite{dash_bitcoin_2014}, Ascribe~\cite{mcconaghy_visibility_2017}, and Plantoid~\cite{okhaos_plantoids_nodate}. NFTs are used to trade in several assets, from collectibles to virtual land; the `art' segment of NFTs has been the leading one so far, with higher average prices and volumes~\cite{nadini_mapping_2021}. In March 2021, NFTs made the news thanks to Beeple’s `Everydays: The First 5000 Days' sale, went for about \$69 million at Christie's. This sale made the artist the third highest-selling living artist at the time. Beyond the hype of large sales, NFTs are by now widely used in a growing ecosystem made of several, small to large marketplaces.

\begin{figure}[h!]
\begin{center}
\includegraphics[width=\linewidth]{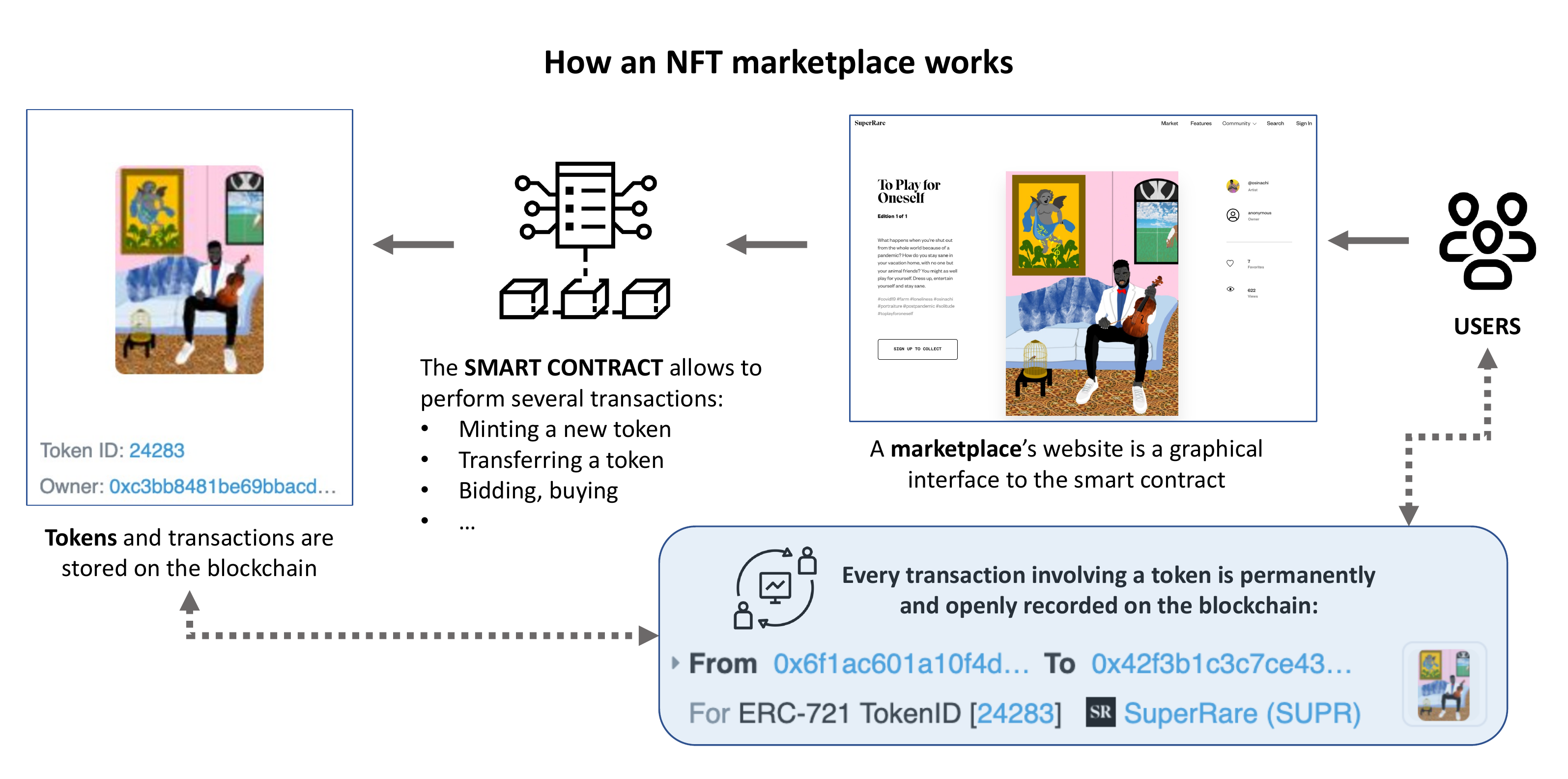}
\end{center}
\caption{Illustration of how an NFT art marketplace works. The marketplace's website is an interface to one or more smart contracts recording events on a blockchain. The example is `\href{https://superrare.com/artwork-v2/to-play-for-oneself-24283}{To Play for Oneself}' by Osinachi, on SuperRare.}\label{fig:1}
\end{figure}

To see how, let us imagine the following scenario, illustrated in Figure~\ref{fig:1}: an artist has created an artwork, either born digital (e.g., digital sculpture or animation) or digitized (e.g., a picture of street art or a video of a performance). This artwork is stored on a computer as a file, therefore it can be endlessly copied. Programmable blockchains like Ethereum allow anyone to record transactions resulting from the interaction with a piece of code, called a `smart contract'~\cite{de_filippi_smart_2021}. By interacting with a suitable smart contract (e.g., via an NFT marketplace such as SuperRare or OpenSea, or by creating their own smart contract), the artist can generate a unique certificate for their artwork. Certificates are called `tokens', and typically contain a reference to the artwork by way of a URL to where it can be accessed online, artwork metadata, or even the input data required to re-create it programmatically (as is the case with generative art, e.g., on ArtBlocks). Thanks to the blockchain, a token is immutably recorded via a transaction, thus creating a transparent, auditable and scarce (i.e., unique or limited-edition) certificate of authenticity for the artwork. The token is at first deposited in the artist's wallet, it can then be exchanged and interacted with, for example traded, used to unlock collector-only rewards and showcased on social media.
 
\subsection{Why many artists and collectors are turning to NFTs}

One reason why NFTs have become rapidly popular in the arts and collectibles markets is frustration with the status quo. The art market supports the economic exchange of creative goods with a primary cultural value. In Bourdieu’s parlance, art is one way to objectify cultural capital into goods~\cite{throsby_cultural_1999}. The main task the art market solves is pricing artworks: a feat not easily accomplished given their heterogeneity and the diverse motivations for buying them~\cite{towse_art_2011,findlay_value_2014}. The art market obtains its supply from living artists, selling from their studio, via galleries and fairs, and from collectors or institutions in all other cases. Intermediaries, such as auction houses, dealers and consultants are active mostly in the secondary market of re-sales~\cite{khaire_culture_2017}. The art market has many peculiarities and outstanding limitations, one of which is that it severely lacks in transparency~\cite{velthuis_contemporary_2012}: some see the presence of information asymmetries as the main cause of limited efficiency in the art market~\cite{david_art_2013,day_explaining_2014}. The attribution, provenance, authenticity, and even copyright of an artwork is often unclear, and this plays a significant role in its ultimate pricing~\cite{cooper_art_2018}. The division between primary and secondary markets creates a further informational barrier, in the context of a general difficulty in estimating the quality of intrinsically unique goods~\cite{gertsberg_asymmetries_2019}, and lacking regulatory frameworks~\cite{doi:10.1177/1023263X221082509}. What is more, price data in the art market is notoriously lacking, hindering the systematic study of artwork prices and pricing over time~\cite{van_miegroet_imperfect_2019}. 

This results is a disparity in the information available to buyers and sellers, placing both at a disadvantage and increasingly allowing prices to be used as signals of quality~\cite{akerlof_market_1970,plattner_most_1998}. Information asymmetries and high uncertainty translate into high transaction costs paid to mediators, they inflate the role of expert judgements and of one's embeddedness in existing networks. In the art market few, powerful and established intermediaries can easily marginalize anyone with reduced access to information~\cite{gertsberg_asymmetries_2019}. It is thus that the reliance on personal and institutional social networks as arbiter for taste and value became commonplace in many cultural and creative sectors, well beyond the fine arts~\cite{godart_how_2009,ebbers_disentangling_2010,williams_quantifying_2019,juhasz_brokering_2020}. Mediators thrive from information asymmetries in the art market~\cite{bonus_credibility_1997,wijnberg_adding_2000,prinz_success_2015,de_silva_market_2016}, and often focus on consensus building among themselves~\cite{kackovic_artists_2020}.

While reducing buyer's uncertainty, gatekeepers channel scarce resources to a limited number of artists, enhancing the superstar phenomenon in the art market~\cite{velthuis_contemporary_2012}. ``Payoffs in these markets are disproportionate to the level of talent and profits are highly rank-dependent. As a result, a small number of individuals absorb the largest portion of revenues''~\cite{gertsberg_asymmetries_2019}. This has a direct consequence on the career prospects of artists. While the working conditions in the art market span a high-variance continuum from tenured contracts to self-employment~\cite{hartog_idiosyncrasies_2019}, ``the overwhelming majority of artists do not enjoy much, if any commercial success''~\cite{plattner_most_1998}. Unpaid or precarious and non-art-related work is not only widespread~\cite{mcrobbie_be_2016,duffy_not_2017}, but also shows stratification by social class, age, and career stage across the cultural and creative industries~\cite{brook_theres_2020}, marginalization~\cite{siebert_all_2013} and alternative copying mechanisms~\cite{merkel_freelance_2019}.

While in recent years the art market has been changing in the direction of increased globalization, financialization, and digitization~\cite{malik_tainted_2012,velthuis_financialization_2012,sidorova_cyber_2019,mcandrew_art_2021}, its underlying market structure has remained static. Even the Internet has had so far limited impact~\cite{deresiewicz_death_2020}, prompting some to say that ``in 2022, [the art market] will look like today, or for that matter, like it did a century and a half ago, when modern art as well as its markets were born''~\cite{velthuis_contemporary_2012}. The rapid success of NFT art must be understood in the context of `static' market structures, operating often without transparency, providing little opportunity to most artists and a significant advantage to gatekeepers.

\subsection{How NFTs might change the art market}

NFTs promise solutions to several challenges~\cite{belsky_furry_2021}: authenticity and provenance are clear and publicly auditable, access to liquidity is easier and immediate, prices are fully transparent, market structures are flexible and customizable, and NFTs are digitally native. The results are already tangible, for example in the drastic reduction of transaction costs~\cite{franceschet_crypto_2021,towse_handbook_2020}. A key opportunity is finding better ways to sustain artistic and creative careers, especially so at their very beginning. NFTs can support social impact investments targeted to artists~\cite{grannemann_artists_2019}, and artist-friendly economic models including fractional equity~\cite{whitaker_artist_2018,whitaker_fractional_2020} and artist royalties~\cite{schten_no_2017,van_haaften-schick_artists_2021}. Indeed in 2021 alone NFT sales volume totals about \$40 billion~\cite{murphy_how_2021}. Compare with the arts and antiques market sized at \$50 billion in 2020, down 22\% on 2019 and 27\% since 2018~\cite{mcandrew_art_2021}.

To be sure, several operational, legal and regulatory questions remain open about the use of NFTs. Many innovations are properties of the NFT, not the artwork itself, and therefore rest to some extent on informal social contracts. The relationship between physical artworks and NFTs is often unspecified, bringing a lack of legal and technical clarity when exchanging NFTs in place of the artworks they relate to. Further open problems include copyright~\cite{savelyev_copyright_2018}, environmental costs~\cite{kay_hidden_2021}, governance~\cite{yermack_corporate_2017}, platform centralization~\cite{marlinspike_my_2022}. Answering or overcoming such challenges would be critical given the adoption and potential of NFTs.

\subsection{What are we missing?}

The discourse around the actual and potential impact of NFTs in the arts is often speculative. On the one hand, critics warn against the risk of furthering the commodification of art and the financialization of artistic practices~\cite{zeilinger_digital_2018}. To some, it is unlikely that blockchain technology will offer an alternative to inequalities in the art world, but it will, instead, reinforce the status quo~\cite{sommerer_economy_2017}. That NFTs are also a speculative financial asset is indisputable, as confirmed by ever abundant work on their profitability, price estimation, and correlation with other asset classes~\cite{ante_non-fungible_2021,dowling_fertile_2021,dowling_is_2021,khezr_property_2021,borri_economics_2022,kraussl_non-fungible_2022,bao_non-fungible_2022}; there are also signs of illicit practices, like wash trading and money laundering~\cite{murphy_how_2021}, although these risks might have been exaggerated~\cite{von_wachter_nft_2022}. Indeed, there are reasons for caution: early empirical studies of the overall NFT market showcase a concentration of opportunity and winner-take-all outcomes not dissimilar to the traditional art world~\cite{nadini_mapping_2021,vasan_quantifying_2022}.

At the same time, it is difficult to dismiss the potential of NFTs affordances to bring about change in the art market, if only given their enthusiastic adoption by creators, the radically lower transaction costs and much improved transparency. Creative designs of new modes of organizing artistic practices and markets via blockchains are just beginning to be explored~\cite{lotti_art_2019}. The companies that have been most successful in the NFTs market make innovative use of the NFT technology (e.g., artist royalties), and leverage the community of users, maintaining community engagement and easy onboarding~\cite{kaczynski_how_2021}. An example is SuperRare, among the first and most successful NFT art marketplaces, which evolved from a privately owned company into a decentralized autonomous organization in Summer 2021, giving partial control to the early community via the \$RARE social token. SuperRare is now jointly governed by its founders, funders and community via proof of stake~\cite{superrare_introduction_2021}. The fact is that NFTs allow for an open design space in terms of market logic, with each marketplace taking different approaches. For example, a marketplace might use fixed artist royalties on secondary sales, or instead allow the artist to adjust them as they prefer. In practice, we do not know what drives the development and adoption of NFTs, which market designs are most successful in fostering both economically and societally beneficial outcomes, which ones have instead a mostly negative impact, and why. 
NFTs are re-enacting the clash between the sphere of culture and the sphere of power in the arts~\cite{bourdieu_field_1993,throsby_production_1994,hanquinet_inequalities_2017}, so that a novel equilibrium might emerge in due course. All evidence from the labor market is telling us that ``our future economy will be built on creativity and technology''~\cite{easton_creativity_2018}. Many cultural and creative professions like journalism, advertising, and communication, have been greatly impacted by participatory media culture~\cite{deuze_convergence_2007,miller_understanding_2020}, big data~\cite{boyd_critical_2012} and algorithms~\cite{mittelstadt_ethics_2016,ananny_seeing_2018,gandy_panoptic_2021}: it might now be the turn of arts.

\section{Data}

Given the focus of this study, we consider a set of galleries for trading unique NFT art. Collectibles and art drops, such as in ArtBlocks, are out of scope. The selected galleries cover the range from large (e.g., SuperRare) to small (e.g., Zora) in terms of sales and their priced volume. We collect data from the OpenSea API,~\footnote{\url{https://docs.opensea.io/reference/api-overview}.} considering all available sales between a gallery's start of operations and August 2022 included.

Table~\ref{tab:dataset_stats} provides an overview of the six galleries and their sales data from OpenSea. Galleries can have a large number of transactions (e.g., Makersplace), a high volume (e.g., Foundation), both (e.g., SuperRare), or neither (e.g., Zora), making for a varied group as intended. This is further evidenced by the average and media sale prices, which vary substantially across galleries. The total dataset comprises almost 40k sales over several years starting in 2018, and a total priced volume of over 237M USD. Every price in this study is given in USD, with the exchange rate calculated at the time of a sale. Lastly, we note that the number of distinct sellers and buyers is small but not negligible in an art market, with 5636 sellers and over 10k buyers. It is important to underline that every seller or buyer is identified by their blockchain addresses: we make no attempt to relate addresses to real-world individuals. Furthermore, not all sellers are artists/creators, and not all buyers are holders/collectors: we also make no attempt to distinguish the primary (first time sales) and secondary (resales) markets in this study. The reason is that non-priced transactions are widespread in NFT art, whereby it is easy to transfer a token from one address to another. In the absence of a reliable way to establish an individual's identity beyond addresses, it follows that distinguishing the primary from secondary market is also not possible in a reliable way.

\begin{table}[]
\caption{Overview of the dataset per gallery. All sale values are in USD converted at the time of the sale.}
\label{tab:dataset_stats}
\begin{tabular}{lrrrrrr}
 &
  \multicolumn{1}{c}{\textbf{Sales}} &
  \multicolumn{1}{c}{\textbf{Volume}} &
  \multicolumn{1}{c}{\textbf{Sale (avg)}} &
  \multicolumn{1}{c}{\textbf{Sale (med)}} &
  \multicolumn{1}{c}{\textbf{Sellers}} &
  \multicolumn{1}{c}{\textbf{Buyers}} \\
SuperRare   & 21,717 & 145,187,662 & 6685   & 457 & 2856 & 4461 \\
Async Art   & 196    & 3,505,909   & 17,887 & 537 & 87   & 143  \\
KnownOrigin & 4306   & 16,344,121  & 3796   & 131 & 1265 & 2412 \\
Foundation  & 1091   & 43,304,375  & 39,692 & 436 & 676  & 782  \\
Makersplace & 11,344 & 28,656,875  & 2526   & 227 & 1220 & 3439 \\
Zora        & 286    & 106,289     & 372    & 104 & 189  & 244  \\
\textbf{Overall} &
  \textbf{38,940} &
  \textbf{237,105,230} &
  \textbf{6089} &
  \textbf{316} &
  \textbf{5636} &
  \textbf{10,011}
\end{tabular}
\end{table}

\section{Results}

In the results section we first briefly provide an overview of the NFT art market as represented by the data we discussed previously. Next, we study the concentration of the market across sellers and buyers. Finally, we explore preferential ties and the relative weight of the top sellers and buyers in one to one relationships.

\subsection{Market overview}

\begin{figure}[h]

\begin{subfigure}{0.5\textwidth}
\includegraphics[width=1.1\linewidth]{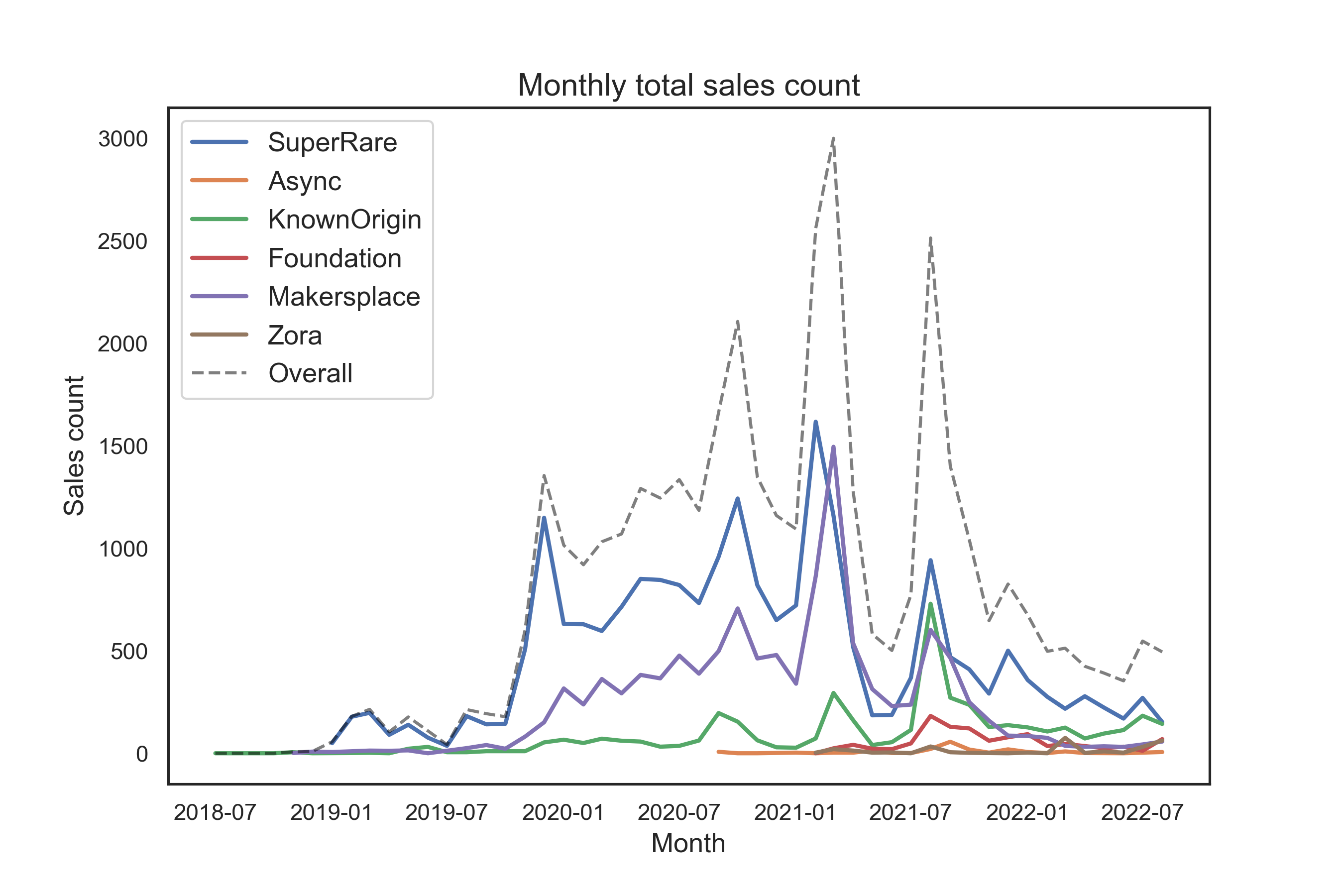} 
\caption{Number of sales.}\label{fig:market_overview_1}
\end{subfigure}
\begin{subfigure}{0.5\textwidth}
\includegraphics[width=1.1\linewidth]{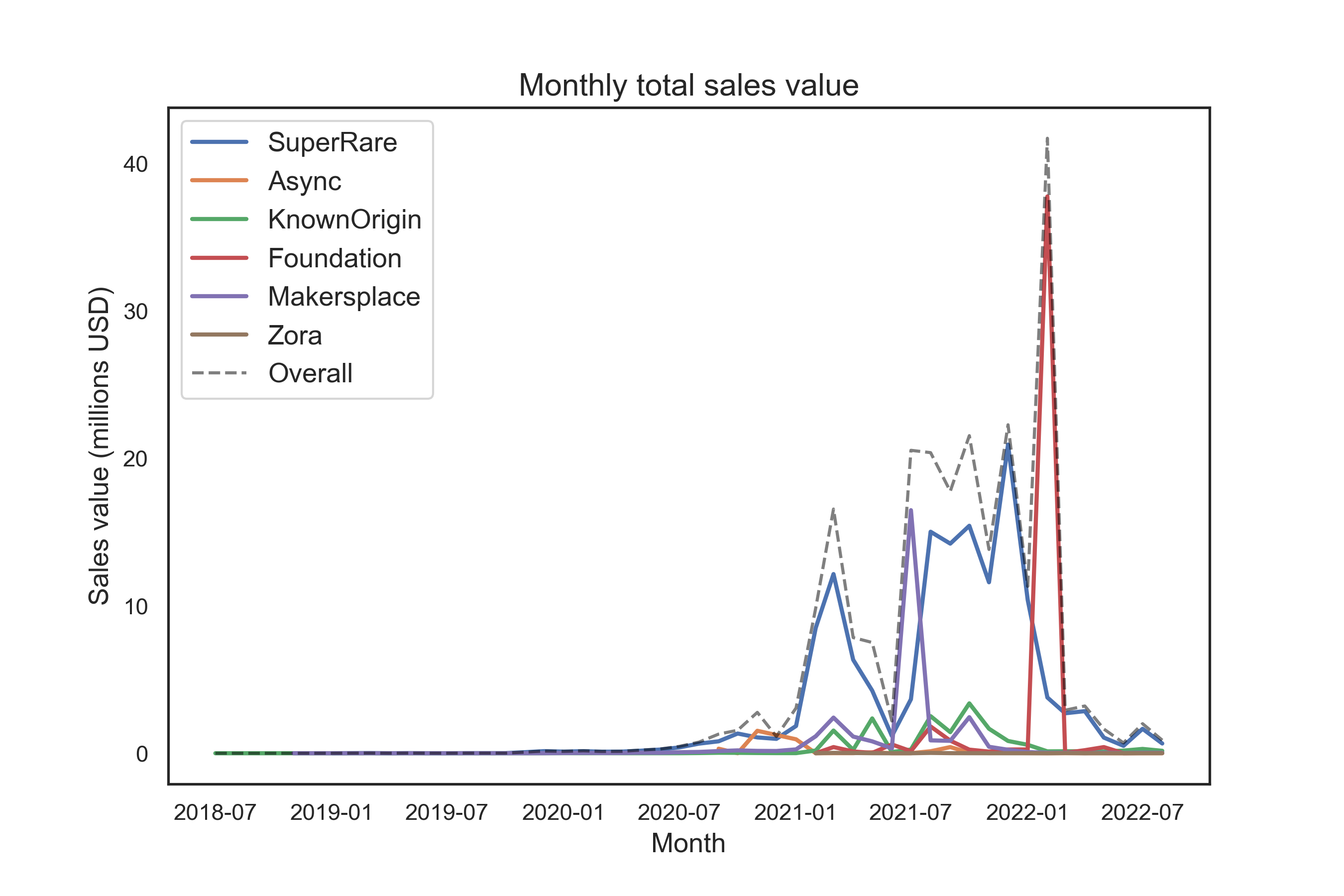}
\caption{Sales volume.}\label{fig:market_overview_2}
\end{subfigure}

\caption{NFT art market overview.}
\label{fig:market_overview}
\end{figure}

Figure~\ref{fig:market_overview} shows an overview of the NFT art market since the start of each gallery, as represented in the dataset, and August 2022 included. Figure~\ref{fig:market_overview_1} provides the monthly number of sales per gallery and overall, while Figure~\ref{fig:market_overview_2} provides the sales volume (in USD). This result clearly shows the trends of NFT art over recent years. While in 2020 the market was active in terms of number of transactions, their volumes remained very low. 2021 has been a boom year for NFT art, and NFTs more generally, even though the market has been driven by the largest galleries (e.g., SuperRare). We are currently in a period of bear market after a significant bust since the beginning of 2022. This study considers a long time period and multiple market cycles with the goal of assessing whether market concentration and preferential ties also change and adapt to the overall market dynamics.

\subsection{Market concentration}

We focus next on market concentration, by investigating to what extent the top sellers and buyers drive the overall market. Firstly, in Table~\ref{tab:market_concentration}, we provide the fraction of sales and priced volume controleld by the top 10\% sellers and buyers, respectively. The overall picture is that of a highly concentrated market, whereby 86\% and 92\% of sales volume is controlled by just the top 10\% of sellers and buyers respectively. The concentration around the number of sales is less pronounced, but still steep. This picture changes somewhat across galleries, with for example the small gallery Zora showing less concentration.  

\begin{table}[]
\caption{Fraction of sales and volume (in USD) controlled by the top 10\% sellers (columns 1 and 2), and by the top 10\% buyers (columns 3 and 4).}
\label{tab:market_concentration}
\begin{tabular}{lrrrr}
                 & \multicolumn{2}{c}{\textbf{Top 10\% sellers}}                            & \multicolumn{2}{c}{\textbf{Top 10\% buyers}}                             \\
                 & \multicolumn{1}{c}{\textit{Sales}} & \multicolumn{1}{c}{\textit{Volume}} & \multicolumn{1}{c}{\textit{Sales}} & \multicolumn{1}{c}{\textit{Volume}} \\
SuperRare   & 67\% & 79\% & 67\% & 88\% \\
Async Art   & 38\% & 84\% & 27\% & 93\% \\
KnownOrigin & 61\% & 86\% & 37\% & 93\% \\
Foundation  & 37\% & 80\% & 30\% & 85\% \\
Makersplace & 78\% & 80\% & 55\% & 76\% \\
Zora        & 33\% & 60\% & 22\% & 68\% \\
\textbf{Overall} & \textbf{71\%}                      & \textbf{86\%}                       & \textbf{64\%}                      & \textbf{92\%}                      
\end{tabular}
\end{table}

To complement these finds we also provide the Lorenz curves showing the concentration of sales and buys. In these Lorenz curves, the sellers and buyers are sorted from top to bottom (in terms of their market activity), and the fraction of sales or volume they control is cumulatively accounted for. In this way, it is easy to show what share of the market a certain fraction of users accounts for.

\begin{figure}[h]

\begin{subfigure}{0.5\textwidth}
\includegraphics[width=1.1\linewidth]{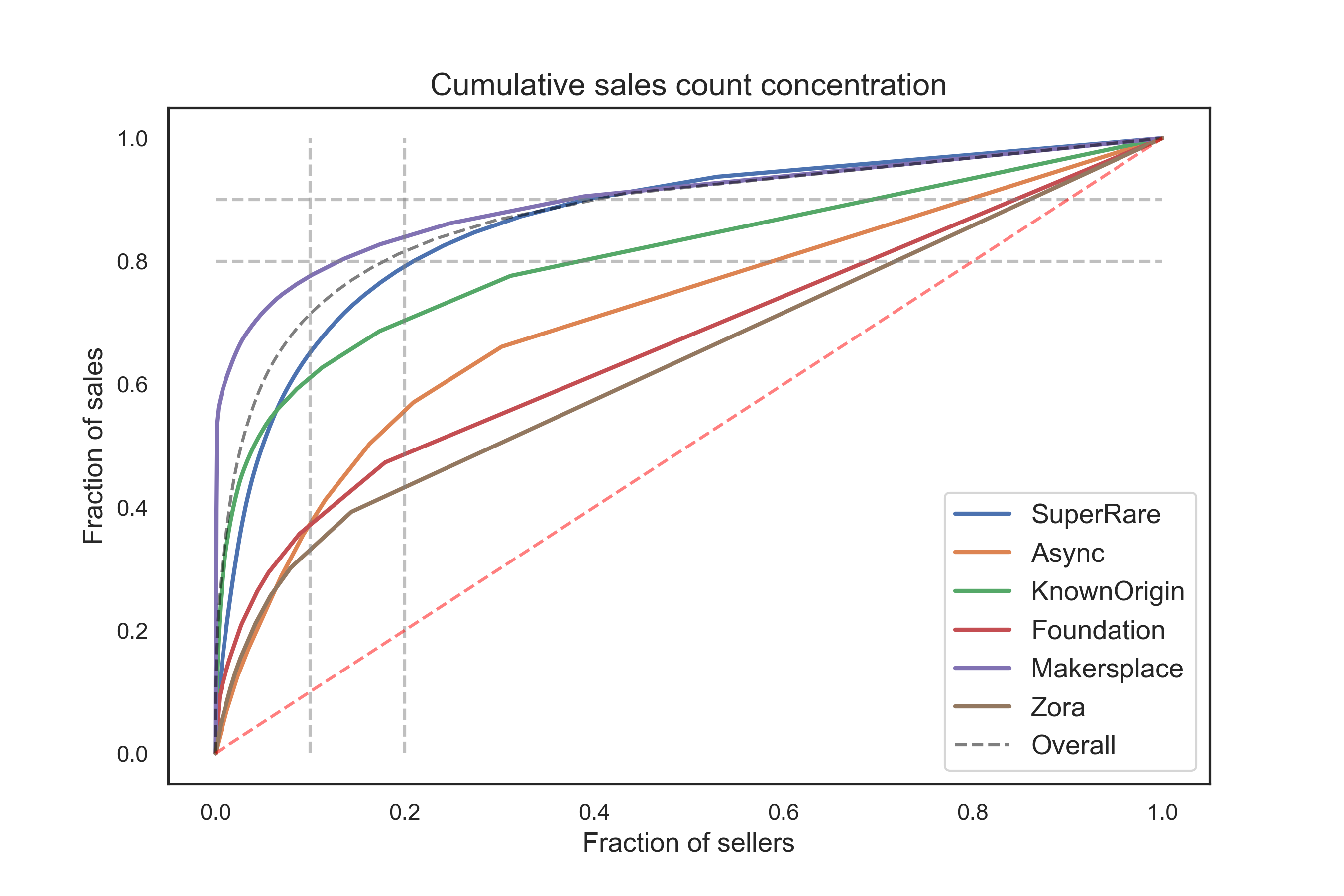} 
\caption{Number of sales.}\label{fig:sales_concentration_1}
\end{subfigure}
\begin{subfigure}{0.5\textwidth}
\includegraphics[width=1.1\linewidth]{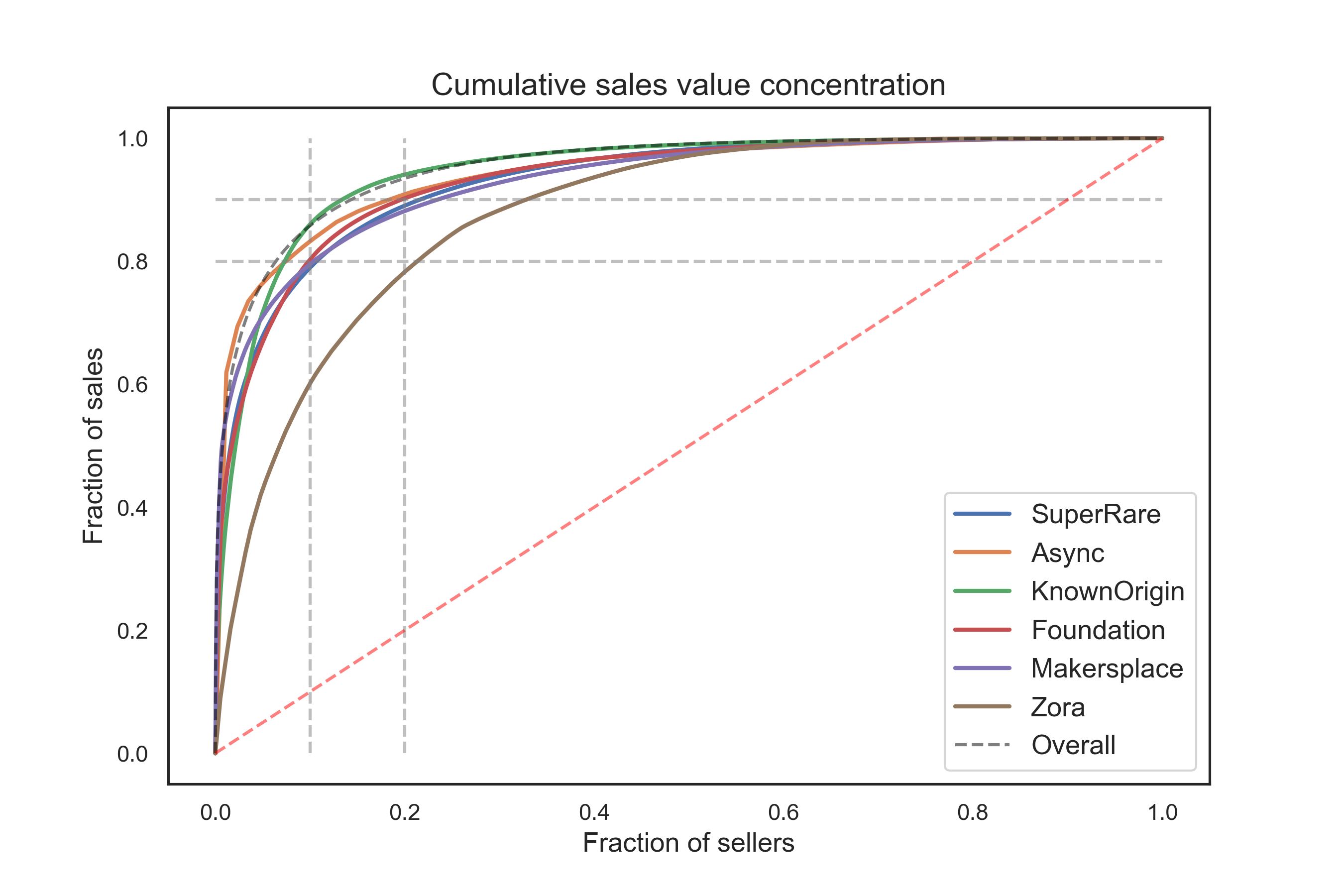}
\caption{Sales volume.}\label{fig:sales_concentration_2}
\end{subfigure}

\caption{NFT art market sales concentration. Black dotted lines provide the 10\%-90\% and the 20\%-80\% intersections, while the red line shows perfect equality.}
\label{fig:sales_concentration}
\end{figure}

Figure~\ref{fig:sales_concentration} show the concentration in the number of sales (Figure~\ref{fig:sales_concentration_1}) and the sales priced volume (Figure~\ref{fig:sales_concentration}). Sellers are sorted on the x axis, while the fraction of sales they control is cumulatively given on the y axis. Black dotted lines can be used to find the 10\%-90\% and 20\%-80\% intersections. The concentration of sale transactions is indeed less pronounced, with Zora, Async and Foundation showing a broader market participation, and SuperRare and Makersplace a narrower one. Yet it is with sales priced volumes that the NFT art market steep concentration clearly shows. For almost all galleries the top 10\% of sellers cover over 80\% of sales volumes, and the top 20\% about a remaining 10\%, with 80\% of the remaining users trading a volume of 10\% of the total, or even less.

\begin{figure}[h]

\begin{subfigure}{0.5\textwidth}
\includegraphics[width=1.1\linewidth]{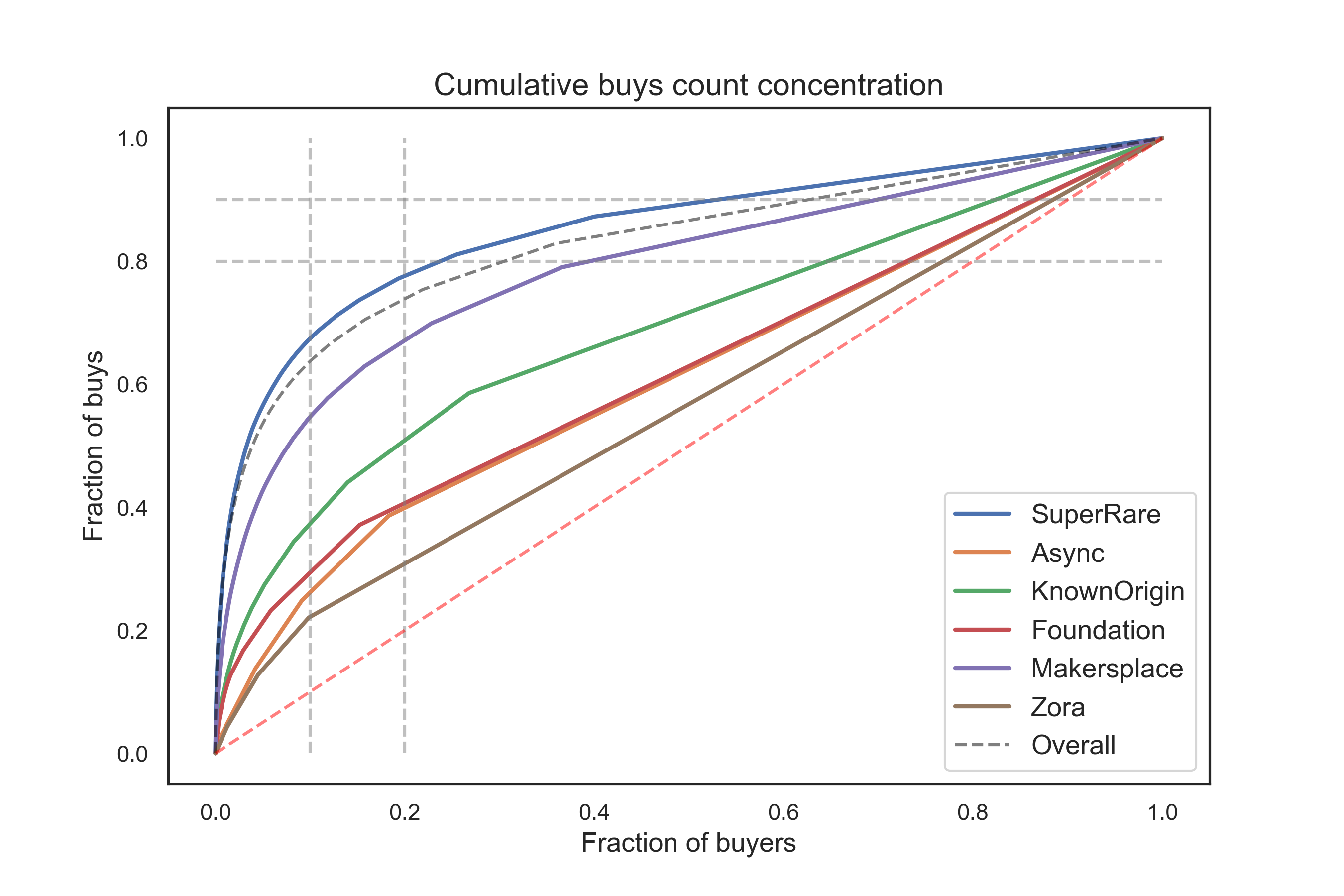} 
\caption{Number of buys.}\label{fig:buys_concentration_1}
\end{subfigure}
\begin{subfigure}{0.5\textwidth}
\includegraphics[width=1.1\linewidth]{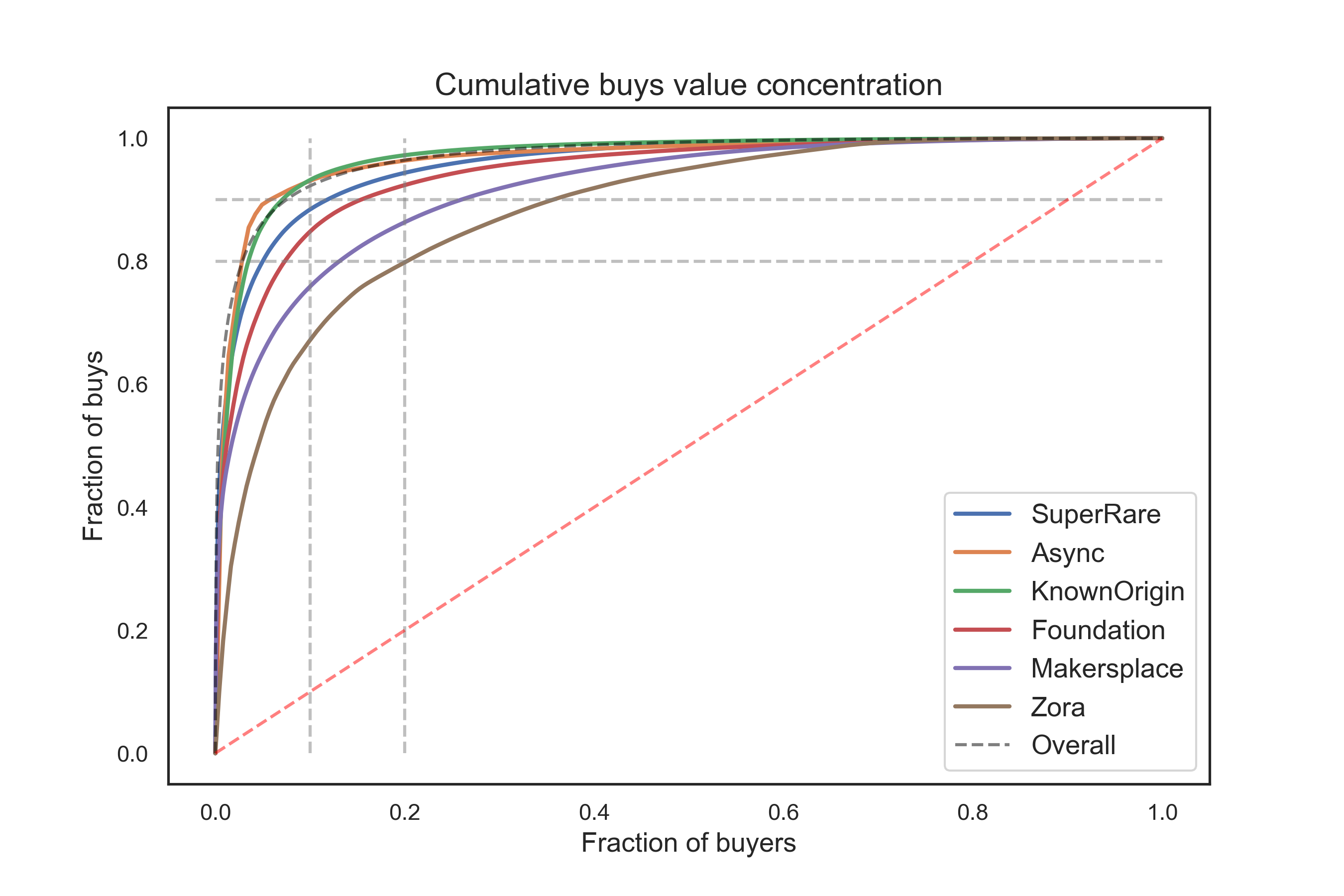}
\caption{Buys volume.}\label{fig:buys_concentration_2}
\end{subfigure}

\caption{NFT art market buys concentration. Black dotted lines provide the 10\%-90\% and the 20\%-80\% intersections, while the red line shows perfect equality.}
\label{fig:buys_concentration}
\end{figure}

We similarly consider buys in Figure~\ref{fig:buys_concentration}, providing Lorenz curves for the number of buys (Figure~\ref{fig:buys_concentration_1}) and their volumes as well (Figure~\ref{fig:buys_concentration_2}). While the concentration of buys is somewhat less pronounced than that of sales, the concentration of buys volumes is even stronger. In galleries such as Async and KnownOrigin, the top 10\% buyers are responsible for over 90\% of buys volume, with SuperRare close behind. Again Zora and, in this case, Makersplace are less concentrated.

Our results on the NFT art market concentration echo previous studies focused on the overall NFT market. \cite{nadini_mapping_2021} find that traders tend to form tight clusters specialized by NFT category. Furthermore, in their study, ``the top 10\% of traders alone perform 85\% of all transactions and trade at least once 97\% of all assets.''

\subsection{Preferential ties}

We operationalize preferential ties as the main seller-buyer directional relationships. Therefore, the preferential tie for a seller is their main buyer, and similarly for a buyer. We then focus on how large a role preferential ties play in the NFT art market. We measure their effect via Equation~\ref{eq:01}.

\begin{equation}
p_a = \frac{v_{a,b}}{v_a}\label{eq:01}
\end{equation}

$p_a$ expresses the preferential tie score of a seller $a$, with their preferential buyer $b$. The preferential buyer is simply defined as the main one by sales or volume. $v_{a,b}$ are sales from $a$ to $b$. $v$ can represent either sales counts or their priced volumes, and $v_a$ is the total number or volume of sales for $a$. $p_a$ thus ranges between zero and one, with one meaning that all buys from $a$ were made by $b$. The same scores can be calculated for buyers and their preferential sellers. This preferential tie score is of immediate interpretation and complements the Lorenz curves discussed above.

In Table~\ref{tab:preferential_ties} we show the average preferential scores across galleries and overall, distinguishing between sellers and buyers. From these results we can immediately appreciate how strongly NFT art is driven by preferential ties: the overall seller to buyer sales are conducted via preferential ties 82\% of times on average, and 86\% for buyer to seller sales. Sale volumes are even more concentrated on preferential ties, with 86\% and 89\% respectively. Interestingly, the gallery relying most on preferential ties is Zora, which emerged as the least market concentrated.

\begin{table}[]
\caption{Average preferential tie scores: the fraction of sales and volume (in USD) from a seller to their preferential buyer (columns 1 and 2), and from a buyer to their preferential seller (columns 3 and 4).}
\label{tab:preferential_ties}
\begin{tabular}{lrrrr}
                 & \multicolumn{2}{c}{\textbf{Seller to buyer}}                             & \multicolumn{2}{c}{\textbf{Buyer to seller}}                             \\
                 & \multicolumn{1}{c}{\textit{Sales}} & \multicolumn{1}{c}{\textit{Volume}} & \multicolumn{1}{c}{\textit{Sales}} & \multicolumn{1}{c}{\textit{Volume}} \\
SuperRare   & 79\% & 83\% & 80\% & 85\% \\
Async Art   & 89\% & 90\% & 96\% & 96\% \\
KnownOrigin & 89\% & 91\% & 90\% & 93\% \\
Foundation  & 94\% & 95\% & 94\% & 96\% \\
Makersplace & 84\% & 87\% & 91\% & 93\% \\
Zora        & 94\% & 94\% & 96\% & 97\% \\
\textbf{Overall} & \textbf{82\%}                      & \textbf{86\%}                       & \textbf{86\%}                      & \textbf{89\%}                      
\end{tabular}
\end{table}

\begin{figure}[h]

\begin{subfigure}{0.5\textwidth}
\includegraphics[width=1.1\linewidth]{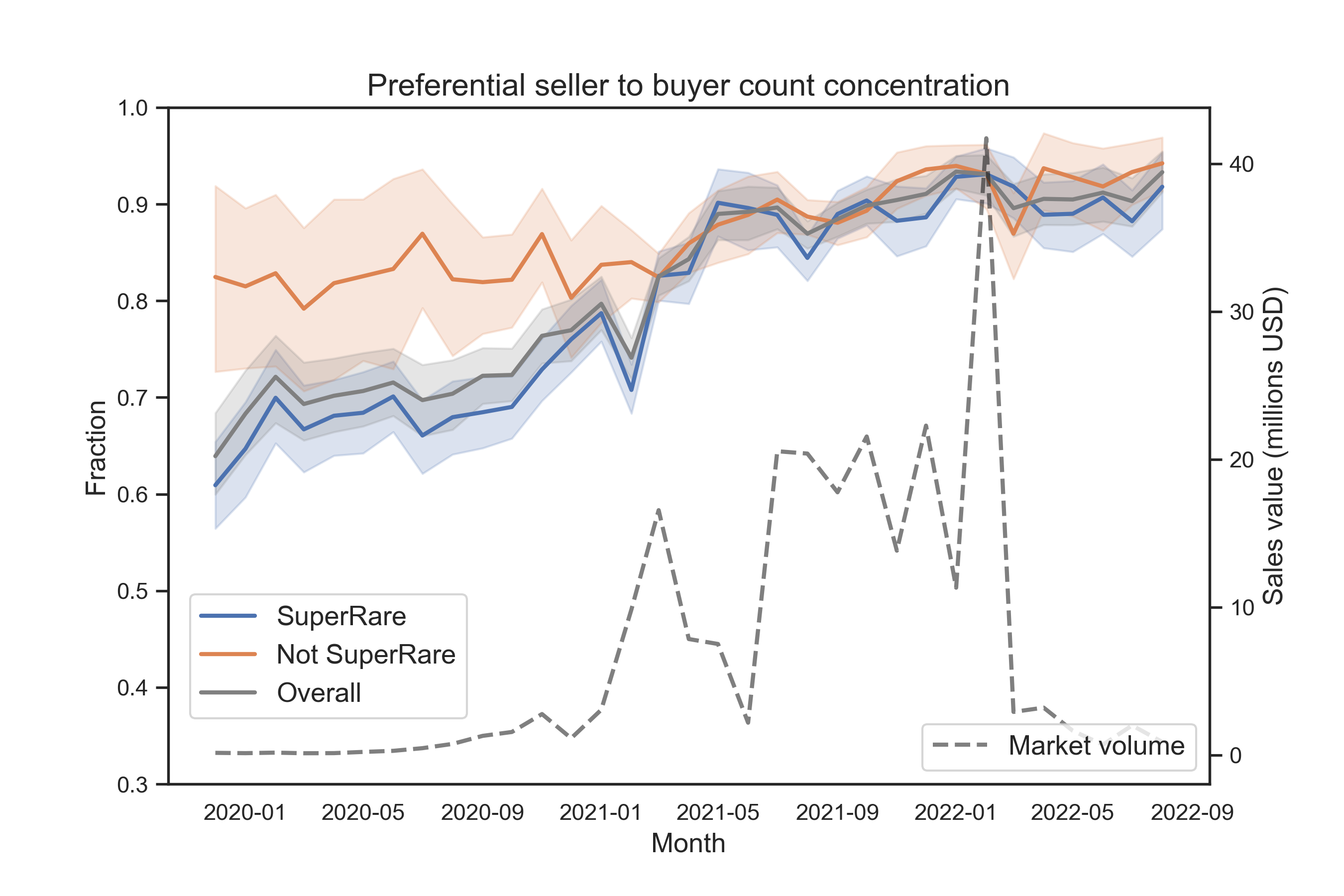}
\caption{Number of sales.}\label{fig:seller_pref_1}
\end{subfigure}
\begin{subfigure}{0.5\textwidth}
\includegraphics[width=1.1\linewidth]{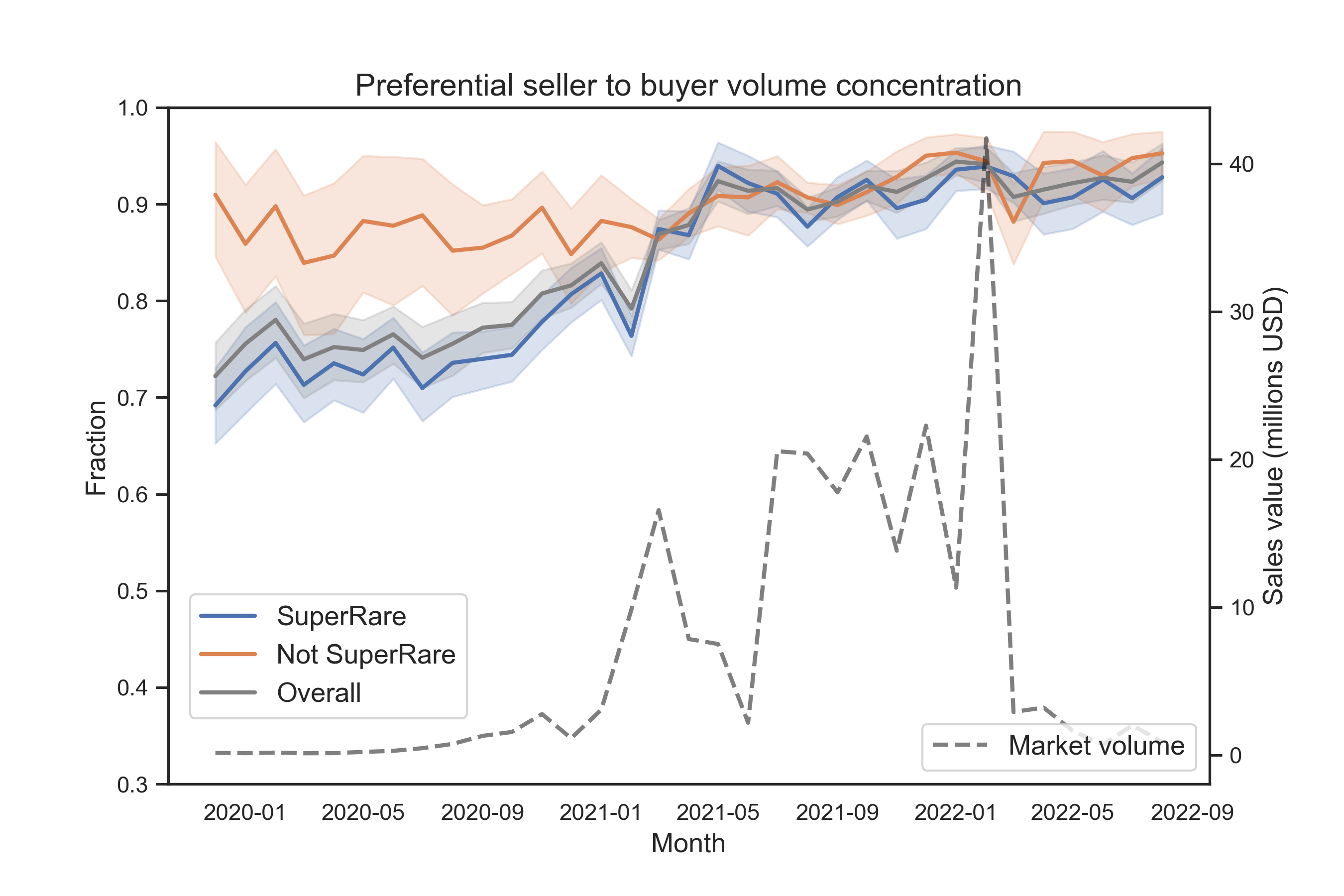}
\caption{Sales volume.}\label{fig:seller_pref_2}
\end{subfigure}

\caption{Seller to buyer preferential ties. The overall market volume, provided for comparison, is the same as the respective line in Figure~\ref{fig:market_overview_2}. 95\% bootstrapped confidence intervals are provided.}
\label{fig:seller_pref}
\end{figure}

\begin{figure}[h]

\begin{subfigure}{0.5\textwidth}
\includegraphics[width=1.1\linewidth]{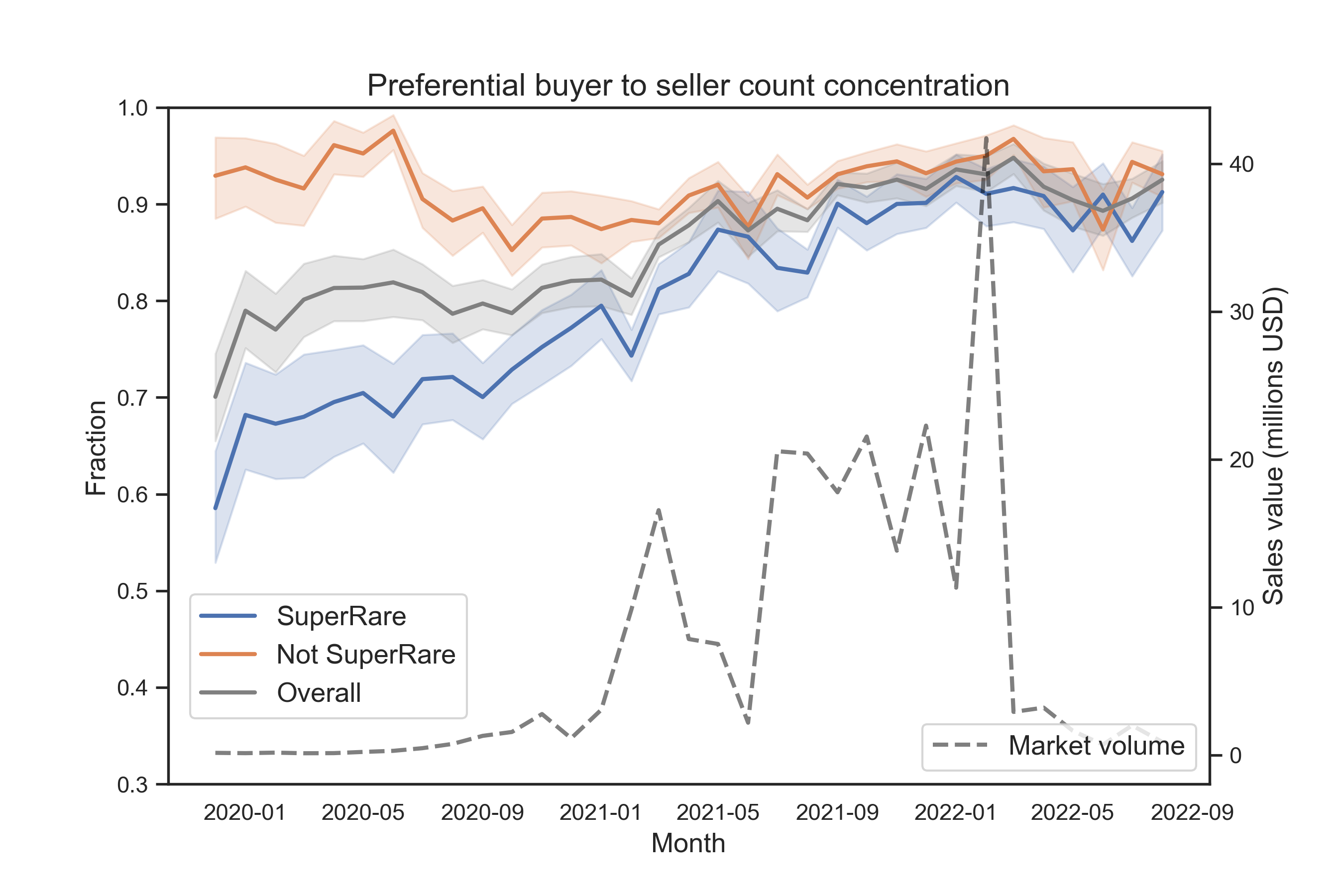}
\caption{Number of sales.}\label{fig:buyer_pref_1}
\end{subfigure}
\begin{subfigure}{0.5\textwidth}
\includegraphics[width=1.1\linewidth]{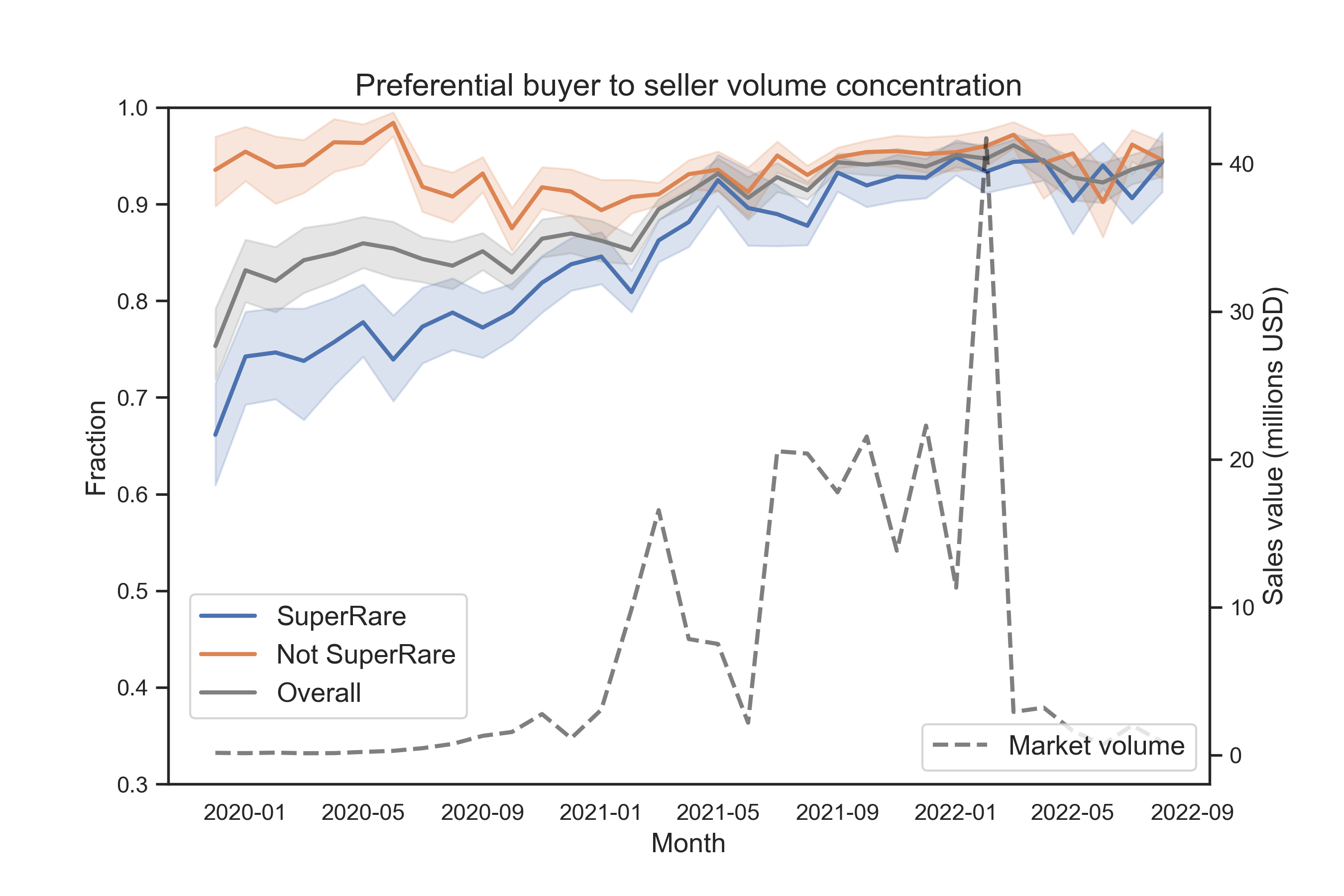}
\caption{Sales volume.}\label{fig:buyer_pref_2}
\end{subfigure}

\caption{Buyer to seller preferential ties. The overall market volume, provided for comparison, is the same as the respective line in Figure~\ref{fig:market_overview_2}. 95\% bootstrapped confidence intervals are provided.}
\label{fig:buyer_pref}
\end{figure}

Next, we show the role of preferential ties over time. Figure~\ref{fig:seller_pref} focuses on seller to buyer ties, while Figure~\ref{fig:buyer_pref} on buyer to seller ties. Both figures show preferential ties for sales (left) and priced volumes (right). These figures show the trend for SuperRare, every other gallery, and overall. This is motivated by the relative size of SuperRare in the dataset, with well over half the sales and volumes overall. Furthermore, both figures highlight the overall market volume over time, for reference. Clear and similar trends emerge for both sellers and buyers. First of all, SuperRare started off as a less concentrated gallery, with preferential ties playing a large yet less dominant role. For the other galleries, this was not the case, with preferential ties always playing a dominant role. The role of preferential ties grew over time in SuperRare, except during the first market boom in the early 2021. We can speculate that the market boom of the 2021, which initially led to a diversification of ties, rapidly consolidated into overly strong preferential ties which have been since dominating the market in all galleries. The 2022 bust does not seem to have made an impact in this respect. Currently, all galleries have average preferential tie scores of over 90\% in all respects.

Our results confirm and expand previous studies. \cite{nadini_mapping_2021} show that ``the distribution of link weights [in NFT markets] is well characterized by a power law distribution, with the top 10\% of buyer–seller pairs contributing to the total number of transactions as much as the remaining 90\%.'' Focusing on the Foundation NFT art gallery, \cite{vasan_quantifying_2022} find that artists receive repeated investments from a small group of leading collectors. This effect is particularly strong for successful artists, who drive the bulk of the market in terms of volume.

\section{Conclusion}

In this work, we consider the recently emerged NFT art market, and investigate its promise to bring about a broader participation and diversity in market participation for both sellers and buyers of art. This contribution started with an effort to position NFT art in the broader context of the art market via an extended overview of previous work. \textbf{The traditional art market is notoriously opaque and hard to access, with steep winner-take-all mechanics making it hard for most artists and collectors to benefit from it. Is the NFT art market any different? The answer is no.} 

We assemble a large dataset of sales across six NFT art galleries, representing different market segments, scales and volumes. Next, we confirm that these galleries follow in the recent history of NFT markets, characterized by a rapid growth in 2020, a boom in 2021, and a bust in 2022. We then consider market concentration and show that, while galleries differ markedly, the top sellers and buyers largely dominate transactions both in terms of sales and priced volume. Lastly, we investigate the role of preferential ties over time, finding that with the boom market all galleries converged on a very high (over 90\%) preferential tie score, which persists in bear market times. Our results confirm and complement recent work at the scale of all the NFT market~\cite{nadini_mapping_2021}, and at the scale of a single NFT art gallery~\cite{vasan_quantifying_2022}.

While our findings are useful to clarify that the NFT art market does not appear to differ from the traditional art market with respect to market concentration and the role of preferential ties, several questions remain open. The main one might be what drives market concentration and preferential ties. A hypothesis is hype and speculation, which is a purview of few wealthy traders, at least at the scale of the NFT market in 2021. A second question concerns whether there are NFT markets which managed to remain less concentrated and broader in participation, and if so how. Such questions constitute compelling ideas for future work.

\section*{Conflict of Interest Statement}

At the time of writing, the author is a co-founder of NiftyValue,~\footnote{\url{https://www.niftyvalue.com}.} a company active in the business of automatic pricing and discovery of NFT art. NiftyValue has not been involved in this study.



\section*{Data Availability Statement}
The data used in this study can be retrieved via OpenSea.~\footnote{\url{https://docs.opensea.io/reference/api-overview}.}

\bibliographystyle{plain}
\bibliography{2022-NFTart}

\end{document}